\documentclass{article}
\usepackage{graphicx}
\usepackage{amsmath,amssymb} 
\newtheorem{theorem}{Theorem}[section]

\newtheorem{definition}{Definition}[section]
\newtheorem{lemma}{Lemma}[section]
\newtheorem{corollary}{Corollary}[section]
\newtheorem{proposition}{Proposition}[section]
\newtheorem{rem}{Remark}[section]

\newtheorem{ass}{Assumptions}

\begin{document}

\title{Space Alternating Penalized Kullback Proximal Point Algorithms for Maximizing Likelihood with Nondifferentiable Penalty
}


\author{St\'ephane Chr\'etien \thanks{Mathematics Department, UMR CNRS 6623 and University of Franche Comte, UFR-ST, 16 route de Gray, 25030 Besan{\c c}on, France.
              {\bf Email}: stephane.chretien@univ-fcomte.fr}        \and
        Alfred Hero \thanks{department of electrical engineering and computer science, The University of Michigan, 1301 Beal Avenue, Ann Arbor,MI 48109-2122, USA. 
{\bf Email}: hero@eecs.umich.edu} \and Herv\'e Perdry \thanks{Universit\'e Paris-Sud and Inserm UMR-S 669, H\^opital Paul Brousse, 94817 Villejuif Cedex, France.
	   {\bf Email}: perdry@vjf.inserm.fr} 
}


%

\maketitle

\begin{abstract}
The EM algorithm is a widely used methodology for penalized likelihood estimation. Provable monotonicity and convergence are the hallmarks of the EM algorithm and these properties are well established for smooth likelihood and smooth penalty functions. However, many relaxed versions of variable selection penalties are not smooth. In this 
paper we introduce a new class of Space Alternating Penalized Kullback Proximal extensions of the EM algorithm for nonsmooth 
likelihood inference. We show that the cluster points of the new method are stationary points even when they lie on 
the boundary of the parameter set.  We illustrate the new class of algorithms for the problems of model selection for finite 
mixtures of regression and of sparse image reconstruction.

\end{abstract}

\section{Introduction}
\label{intro}

The EM algorithm of Dempster Laird and Rudin (1977) is a widely applicable methodology for computing likelihood 
maximizers or at least stationary points. It has been extensively studied over the years and many useful generalizations have been proposed including, for instance, the stochastic EM algorithm of Delyon, Lavielle and Moulines (1999) and Kuhn and Lavielle (2004); the PX-EM accelerations of Liu, Rubin and Wu (1998); 
the MM generalization of Lange and Hunter (2004) and approaches using extrapolation such as proposed in Varadhan and Roland (2007).

In recent years, much attention has been given to the problem of variable selection for multiparameter estimation, for which the desired solution is sparse, i.e. 
many of the parameters are zero. Several approaches have been proposed for recovering sparse models. A large number of 
contributions are based on the use of non-differentiable penalties like the LASSO (Tibshirani (1996) and Cand\` es and Plan (2008)), ISLE (Friedman and Popescu (2003)) and "hidden variable"-type approach developed by Figueiredo and Nowak (2003). Other contributions are
for instance sparse Bayes learning (Tipping (2001)),  information theoretic based prior methods of Barron (1999), empirical Bayes (Johnstone and Silverman (2004)). Among recent alternatives is 
the new Dantzig selector of Cand\`es and Tao (2008). On the other hand, only a few attempts have been made to use of non-differentiable
penalization for more complex models than the linear model; for some recent progress, see Koh, Kim, and Boyd (2007) for the case of logistic regression; and Khalili and Chen (2007) for mixture models. 

In the present paper, we develop new extensions of the EM algorithm that incorporate a non-differentiable 
penalty at each step. Following previous work of the first two authors, we use a Kullback Proximal interpretation for 
the EM-iterations and prove stationarity of the cluster points of the methods using nonsmooth analysis tools.
Our analysis covers coordinate by coordinate methods such as Space Alternating extensions of EM and Kullback Proximal 
Point (KPP) methods. Such component-wise versions of EM-type algorithms can benefit from acceleration of convergence speed (Fessler and Hero (1994)). The KPP method was applied to gaussian mixture models
in Celeux {\em et al.} (2001). The main result of this paper is that any cluster point of the Space Alternating 
KPP method satisfies a nonsmooth Karush-Kuhn-Tucker condition.

The paper is organized as follows. In section 2 we review Penalized Kullback 
Proximal Point methods and introduce componentwise PKPP algorithms with new differentiable penalties. 
In Section 3, our main asymptotic results are presented. In Section 4, we present a 
space alternating implementation of the penalized EM algorithm for a problem of model selection 
in a finite mixture of linear regressions using the SCAD penalty introduced in Fan and Li (2001) and further studied in Khalili and Chen (2007).

\section{The EM algorithm and its Kullback proximal generalizations}
The problem of maximum likelihood (ML) estimation consists of solving the maximization
\begin{equation}
\label{ml}
\theta_{ML} = {\rm argmax}_{\theta \in \Theta} \; l_y(\theta),
\end{equation}
where $y$ is an observed sample of a random variable $Y$ defined on a
sample space $\mathcal Y$ and $l_y(\theta)$ is the log-likelihood
function defined by
\begin{equation}
l_y(\theta)=\log g(y;\theta),
\nonumber
\end{equation}
on the parameter space $\Theta\subset \mathbb R^p$,
and $g(y;\theta)$ denotes the density of $Y$ at $y$ parametrized by the vector
parameter $\theta$.

The standard EM approach to likelihood maximization introduces a complete data vector 
$X$ with density $f$. Consider the conditional density function $k(x| y;\bar{\theta})$ of $X$ given $y$
\begin{equation}\label{condi}
k(x| y;\bar{\theta})=\frac{f(x;\bar{\theta})}{g(y;\bar{\theta})}.
\end{equation}
As is well known, the EM algorithm then consists of alternating between two steps. The first step, called the E(xpectation) step, 
consists of computing the conditional expectation of the complete log-likelihood given $Y$. Notice that the conditional 
density $k$ is parametrized by the current iterate of the unknown parameter value, denoted here by $\bar{\theta}$ for simplicity.
Moreover, the expected complete log-likelihood is a function of the variable $\theta$. Thus the second step, called the 
M(aximization) step, consists of maximizing the obtained expected complete log-likelihood with respect to the variable  
parameter $\theta$. The maximizer is then accepted as the new current iterate of the EM algorithm and the two 
steps are repeated until convergence is achieved. 

Consider now the general problem of maximizing a concave function $\Phi(\theta)$.
The original proximal point algorithm introduced by Martinet (1970)
is an iterative procedure which can be written
\begin{equation}
\label{prox}
\theta^{k+1}={\rm argmax}_{\theta \in D_\Phi}\left\{\Phi(\theta)
-\frac{\beta_k}{2} \|\theta-\theta^k\|^2\right\}.
\end{equation}
The influence of the quadratic penalty $\frac12 \|\theta-\theta^k\|^2$ is controlled by the
sequence of positive parameters $\{\beta_k\}$. Rockafellar (1976) showed that superlinear
convergence of this method occurs when the sequence $\{\beta_k\}$
converges to zero. A relationship between Proximal Point algorithms and EM algorithms 
was discovered in Chr\'etien and Hero (2000) (see also Chr\'etien and Hero (2008) for details). We review the EM analogy to KPP methods
to motivate the space alternating generalization. 
Assume that the family of conditional densities $\{k(x| y;\theta)\}_{\theta \in {\mathbb R}^p}$ is regular in the sense of Ibragimov
and Khasminskii (1981), in particular 
$k(x|y;\theta)\mu(x)$ and $k(x| y;\bar{\theta)}\mu(x)$ are mutually absolutely
continuous for any $\theta$ and $\bar{\theta}$ in ${\mathbb R}^p$. Then the
Radon-Nikodym derivative $\frac{k(x| y,\bar{\theta})}{k(x| y;\theta)}$ exists
for all $\theta,\bar{\theta}$ and we can define the following Kullback Leibler divergence:
\begin{equation}
\label{kullb}
I_y(\theta,\bar{\theta})={\sf E}\bigl[
\log \frac{k(x| y,\bar{\theta})}{k(x| y;\theta)}| y;\bar{\theta} \;
\bigr].
\end{equation}
Let us define $D_l$ as the domain of $l_y$, $D_{I,\theta}$ the domain of $I_y(\cdot,\theta)$ and
$D_I$ the domain of $I_y(\cdot,\cdot)$. Using the distance-like function $I_y$, the Kullback Proximal Point algorithm 
is defined by 
\begin{equation}
\label{proxit}
\theta^{k+1}={\rm argmax}_{\theta \in D_\Phi}\left\{\Phi(\theta)
-\beta_k I_y(\theta,\bar{\theta})\right\}.
\end{equation}
The following was proved in Chr\'etien and Hero (2000).
\begin{proposition}{\rm [Chr\'etien and Hero (2000) Proposition 1].}
\label{equiem}
In the case where $\Phi$ is the log-likelihood, the EM algorithm is a special instance of the Kullback-proximal algorithm with 
$\Phi$ equal to the penalized log-likelihood and $\beta_k =1$, for all $k\in \mathbb N$.
\end{proposition}

\subsection{The Space Alternating Penalized Kullback-Proximal method}

In what follows, and in anticipation of component-wise implementations of penalized KPP, we will use the notation 
$\Theta_r(\theta)$ for the local decomposition at $\theta$ defined by $\Theta_r(\theta)=\Theta \cap \left(\theta +\mathcal S_r\right)$, $r=1,\ldots,R$ 
where $\mathcal S_1,\ldots,\mathcal S_R$ are subspaces of $\mathbb R^p$ and $\mathbb R^p=\oplus_{r=1}^R \mathcal S_r$. 

Then, the Space Alternating Penalized Proximal Point Algorithm is defined as follows.
\begin{definition}
\label{SpAlt}
Let $\psi$: $\mathbb R^p \mapsto \mathcal S_1\times \cdots \times \mathcal S_R$ be a continuously differentiable mapping and let $\psi_r$ 
denote its $r^{th}$ coordinate. 
Let $(\beta_k)_{k\in \mathbb N}$ be a sequence of positive real numbers and $\lambda$ be a positive real vector in $\mathbb R^R$.
Let $p_n$ be a nonnegative possibly nonsmooth locally Lipschitz penalty function with bounded Clarke-subdifferential (see the Appendix for details) on compact sets. 
Then, the Space Alternating Penalized Kullback Proximal Algorithm is defined by
\begin{equation}
\label{kullprox}
\theta^{k+1}={\rm argmax}_{\theta\in \Theta_{k-1 ({\rm mod}\:\: R)+1}(\theta^k)\cap D_l\cap D_{I,\theta^k}}
\left\{l_y(\theta)-\sum_{r=1}^R\lambda_r p_n(\psi_r(\theta))- \beta_k I_y(\theta,\theta^k)\right\},
\end{equation}
where $D_l$ is the domain of $l_y$ and $D_{I,\theta}$ is the domain of $I_y(\cdot,\theta)$.
\end{definition}

The standard Kullback-Proximal Point algorithms as defined in Chr\'etien and Hero (2008) is obtained as special case by selecting 
$R=1$, $\Theta_1=\Theta$, $\lambda=0$. 

The mappings $\psi_r$ will simply be the projection onto the subspace $\Theta_r$, $r=1,\ldots,R$ in the sequel but the proofs 
below allow for more general mappings too. 

\subsection{Notations and assumptions}

The notation $\|\cdot\|$ will be used to denote the norm on any
previously defined space. The space on
which the norm operates should be obvious from the context. For any bivariate
function $\Phi$, $\nabla_1\Phi$ will denote the gradient with respect to the first variable.
For the convergence analysis, we will make the following assumptions. For a locally Lipschitz function $f$, $\partial f(x)$ 
denotes the Clarke subdifferential of $f$ at $x$ (see the Appendix). Regular locally Lipschitz functions are defined 
in the Appendix. 
\begin{ass}
\label{ass1}
{\rm (i) $l_y$ is differentiable and $l_y(\theta)-\sum_{r=1}^R\lambda_r p_n(\psi_r(\theta))$ converges to $-\infty$ whenever $\|\theta\|$ tends to $+\infty$. 
The function $p_n$ is locally Lipschitz and regular.  \\
(ii) The domain $D_{I,\theta}$ of $I(\cdot,\theta)$ is a subset of the domain $D_l$ of $l$. \\
(iii) $(\beta_k)_{k\in\mathbb N}$ is a convergent nonnegative sequence of real numbers whose
limit is denoted by $\beta^*$.\\
(iv)  The mappings $\psi_r$ are such that 
\begin{equation}
\psi_r(\theta+\epsilon d)=\psi_r(\theta) 
\nonumber
\end{equation}
for all $\theta$ in $\Theta$, all $d\in \mathcal S_r^\perp$ and $\epsilon>0$ sufficiently small so that $\theta+\epsilon d
\in \Theta$, $r=1,\ldots,R$. This condition is satisfied for linear projection operators.}
\end{ass}
We will also impose  one of the two following sets of assumptions on the distance-like function $I_y$ in (\ref{kullb}).
\begin{ass}
\label{ass2} {\rm (i) There exists a finite dimensional euclidean
space $S$, a differentiable mapping $t : D_l \mapsto S$ and a
functional $\Psi : D_\Psi \subset S\times S \mapsto \mathbb R$
such that KL divergence (\ref{kullb}) satisfies
$$
I_y(\theta,\bar{\theta})=\Psi(t(\theta),t(\bar{\theta})),
$$
where $D_\psi$ denotes the domain of $\Psi$. \\
(ii) For any $\{(t^k,t)_{k\in \mathbb N}\}\subset D_\Psi$ there
exists $\rho_t>0$ such that $\lim_{\|t^k-t\|\rightarrow\infty}
I_y(t^k,t)\geq \rho_t$.
Moreover, we assume that $\inf_{t\in M}\rho_t>0$ for any bounded set $M\subset S$.  \\
For all $(t^\prime,t)$ in $D_\Psi$, we will also require that \\
(iii) (Positivity) $\Psi(t^\prime,t)\geq 0$, \\
(iv) (Identifiability) $\Psi(t^{\prime},t)=0 \Leftrightarrow t=t^\prime$,\\
(v) (Continuity) $\Psi$ is continuous at $(t^{\prime},t)$ \\
and for all $t$ belonging to the projection of $D_\Psi$ onto its
second coordinate,\\
(vi) (Differentiability) the function $\Psi(\cdot,t)$ is
differentiable at $t$.}
\end{ass}

In the case where the Kullback divergence $I_y$ is not defined everywhere (for instance if its
domain of definition is the positive orthant), we need stronger assumptions to  
prove the desired convergence properties. 
\begin{ass}
\label{ass3} {\rm (i) There exists a differentiable mapping $t : D_l \mapsto \mathbb R^{n\times m}$ such that 
the Kullback distance-like function $I_y$ is of the form
$$
I_y(\theta,\bar{\theta})=\sum_{1\leq i\leq n, 1\leq j\leq m} \alpha_{ij}(y_j)t_{ij}(\theta)
\phi\Big(\frac{t_{ij}(\bar{\theta})}{t_{ij}(\theta)}\Big),
$$
where for all $i$ and $j$, $t_{ij}$ is continuously differentiable on its domain of definition,
$\alpha_{ij}$ is a function from $\mathcal Y$ to $\mathbb R_+$, the set of positive real
numbers, \\
(ii) The function $\phi$ is a non negative differentiable convex function defined 
$\mathbb R_*^+$ and such that $\phi(\tau)=0$ if and only if $\tau=1$.\\
(iii) There exists $\rho>0$ such that 
$$\lim_{\mathbb R_+\ni \tau\rightarrow\infty} \phi(\tau)\geq \rho.$$\\
(iv) The mapping $t$ is injective on each $\Theta_r$.
}
\end{ass}

In the context of Assumptions \ref{ass3}, $D_I$ is simply the set
$$
D_I=\{\theta \in \mathbb R^p \mid t_{ij}(\theta)>0 \hspace{.3cm} \forall i\in \{1,\ldots,n\} \text{ and } j
\in \{1,\ldots,m\}  \}^2.
$$
Notice that if $t_{ij}(\theta)=\theta_i$ and $\alpha_{ij}=1$ for all $i$ and all $j$, the functions
$I_y$ turn out to reduce to the well known $\phi$ divergence defined in Csisz\`ar (1967).
Assumptions \ref{ass3} are satisfied by most standard examples (for instance Gaussian 
mixtures and Poisson inverse problems) with the choice $\phi(\tau)=\tau\log(\tau)-1$. 

Assumptions \ref{ass1}(i) and (ii) on $l_y$ are standard and are easily checked
in practical examples, e.g. they are satisfied for the Poisson and additive mixture models. 

Finally we make the following general assumption.

\begin{ass}
\label{ass4}
The Kullback proximal iteration (\ref{kullprox}) is well
defined, i.e. there exists at least one maximizer of (\ref{kullprox}) at each iteration $k$.
\end{ass}
In the EM case, i.e. $\beta=1$, this last assumption is equivalent to the 
computability of M-steps. In practice it suffices to show the inclusion 
$0 \in\nabla l_y(\theta)-\lambda \partial p_n(\psi(\theta))- \beta_k \nabla I_y(\theta,\theta^k)$ for $\theta=\theta^{k+1}$
in order to prove that the solution is unique. Then assumption \ref{ass1}(i) is sufficient for a maximizer to exist.

These technical assumptions play an important role in the theory developed below. 
Assumption 1 (i) on differentiability of the log-likelihood
is important for establishing the Karush-Kuhn-Tucker optimality conditions for cluster points. 
The fact that the objective should decrease to negative infinity as 
the norm of the parameter goes to infinity is often satisfied, or can be easily imposed,
and is used later to garantee boundedness of the sequence of 
iterates. The fact that $p_n$ is regular is standard since the usual
choices are the $\ell_1$-norm, the $\ell_p$-quasi-norms for $0<p<1$, the SCAD penalty, etc ... 
Assumption 1 (ii) is only needed in order to simplify the analysis since, otherwise, 
each iterate would lie in the intersection of $D_l$ and $D_I$ 
and this would lead to asymptotic complications; this assumption is always satisfied in the models we have 
encountered in practice. Assumption 1 (iii) is standard.
Assumption 1 (iv) is satisfied when $\psi_r$ is a projection onto $\mathcal S_r$ 
and simplifies the proofs. Assumption 2 
imposes natural conditions on the "distance" $I_y$. Assumption 2 (ii) ensures that the "distance" $I_y$ is large between 
points whose euclidean distance goes to $+\infty$, thus weakening the assumption that $I_y$ should grow to $+\infty$ in such a case. Assumptions 3 are 
used to obtain the Karush-Kuhn-Tucker conditions in Theorem 2. For this Theorem, we require $I_y$ to behave like a standard Kullback-Leibler "distance" and therefore that $I_y$ has a more constrained shape. 
Assumption 3 (iii) is a simplification of Assumption 2 (ii). Assumption 3 (iv) is a natural injectivity 
requirement.

\section{Asymptotic properties of the Kullback-Proximal iterations}
\label{theo}

\subsection{Basic properties of the penalized Kullback proximal algorithm}

Under Assumptions \ref{ass1}, we state basic properties of the penalized Kullback Proximal Point Algorithm. 
The most basic property is the monotonicity of the 
penalized likelihood function and the boundedness of the
penalized proximal sequence $(\theta^k)_{k\in \mathbb N}$. 
The proofs of the following lemmas are given, for instance, in Chr\'etien and Hero (2000) for the unpenalized 
case ($\lambda=0$) and their generalizations to the present context is straightforward.

We start with the following monotonicity result.
\begin{lemma}
\label{truit} For any iteration $k \in \mathbb N$, the sequence $(\theta^k)_{k\in \mathbb N}$ satisfies
\begin{equation}
\label{majgrad} l_y(\theta^{k+1})-\sum_{r=1}^R \lambda_r p_n(\psi_r(\theta^{k+1}))-(l_y(\theta^k)-\sum_{r=1}^R\lambda_r p_n(\psi_r(\theta^k)))\geq \beta_k I_y(\theta^k,\theta^{k+1})\geq 0.
\end{equation}
\end{lemma}

\begin{lemma}
\label{boundu}
The sequence $(\theta^k)_{k\in \mathbb N}$ is bounded.
\end{lemma}

The next lemma will also be useful and its proof in the unpenalized case where $\lambda=0$ is given in Chr\'etien and Hero (2008) Lemma 2.4.3. The 
generalization to $\lambda>0$ is also straightforward.
\begin{lemma}
\label{yal} Assume that in the Space Alternating KPP sequence $(\theta^k)_{k\in \mathbb N}$, there exists a subsequence
$(\theta^{\sigma(k)})_{k\in \mathbb N}$ belonging to a compact set
$C$ included in $D_l$. Then,
\begin{equation}
\lim_{k\rightarrow\infty} \beta_{k}
I_y(\theta^{k+1},\theta^{k})=0.
\nonumber
\end{equation}
\end{lemma}
One important property, which is satisfied in practice, is that the distance 
between two successive iterates decreases to zero. This property is critical to 
the definition of a stopping rule for the algorithm. 
This property was established in Chr\'etien and Hero (2008) in the case $\lambda =0$. 
\begin{proposition}{\rm [Chr\'etien and Hero (2008) Proposition 4.1.2]}
\label{nondegphi}
The following statements hold.

(i) For any sequence $(\theta^k)_{k\in \mathbb N}$ in $\mathbb R_+^p$ and any bounded sequence
$(\eta^k)_{k\in \mathbb N}$ in $\mathbb R_+^p$, if 
$\lim_{k\rightarrow +\infty} I_y(\eta^k,\theta^k)=0$ then
$\lim_{k\rightarrow +\infty} |t_{ij}(\eta^k)-t_{ij}(\theta^k)|=0$ for all $i$,$j$
such that $\alpha_{ij}\neq 0$.

(ii) If $\lim_{k\rightarrow +\infty} I_y(\eta^k,\theta^k)=0$ and one coordinate of one of the two sequences $(\theta^k)_{k\in \mathbb N}$ and $(\eta^k)_{k\in \mathbb N}$ tends to infinity, so does the other's same coordinate.
\end{proposition}

\subsection{Properties of cluster points}

The results of this subsection state that any cluster point $\theta^*$ such that
$(\theta^*,\theta^*)$ lies on the closure of $D_I$ satisfies a modified Karush-Kuhn-Tucker type 
condition. 
We first establish this result in the case where Assumptions \ref{ass2} hold in addition to Assumptions \ref{ass1} and \ref{ass2} 
for the Kullback distance-like function $I_y$. 

For notational convenience, we define 
\begin{equation}\label{regfunc}
F_\beta (\theta,\bar{\theta})=l_y(\theta)-\sum_{r=1}^R\lambda_r  p_n(\psi_r(\theta))-\beta
I_y(\theta,\bar{\theta}).
\end{equation}

\begin{theorem}
\label{bord}
Assume that Assumptions \ref{ass1}, \ref{ass2} and \ref{ass4} hold and if $R>1$, then, for each $r=1,\ldots,R$, 
$t$ is injective on $\Theta_r$. Assume that the limit of $(\beta_k)_{k\in \mathbb N}$, $\beta^*$, is positive. 
Let $\theta^*$ be a cluster point of the Space Alternating Penalized Kullback-proximal sequence (\ref{kullprox}). 
Assume the mapping $t$ is differentiable at $\theta^*$. If
$\theta^*$ lies in the interior of $D_l$, then $\theta^*$ is a stationary point of the penalized log-likelihod function $l_y(\theta)$, i.e.
$$
0\in \nabla l_y(\theta^*)-\sum_{r=1}^R\lambda_r \partial p_n(\psi_r(\theta^*)).
$$
\end{theorem}
{\bf Proof}. We consider two cases, namely the case where $R=1$ and the case where $R>1$. 

A. If $R=1$ the proof is analogous to the proof of Theorem 3.2.1 in Chr\'etien and Hero (2008). In particular, we have  
\begin{equation}
\label{A}
F_{\beta^*}(\theta^*,\theta^*)\geq F_{\beta^*}(\theta,\theta^*) 
\nonumber
\end{equation}
for all $\theta$ such that $(\theta,\theta^*)\in D_I$. Since $I_y(\theta,\theta^*)$ is differentiable at $\theta^*$, the result 
follows by writing the first order optimality condition at $\theta^*$ in (\ref{A}). 

B. Assume that $R>1$ and let $(x^{\sigma(k)})_{k\in \mathbb N}$ be a subsequence of iterates of (\ref{kullprox}) converging to $\theta^*$. Moreover let $r=1,\ldots,R$ and 
$\theta \in \Theta_r\cap D_l$. For each $k$, let $\sigma_r(k)$ the smallest index
greater than $\sigma(k)$, of the form $\sigma(k^\prime)-1$, with $k^\prime \in \mathbb N$ and $(\sigma(k^\prime)-1) \:({\rm mod}\: R)+1=r$. 
Using the fact that $t$ is injective on every $\Theta_r$, $r=1,\ldots,R$, Lemma \ref{yal} and the fact that $(\beta_k)_{k\in \mathbb N}$ converges to $\beta^*>0$, we easily conclude that $(\theta^{\sigma_r(k)})_{k\in \mathbb N}$ and $(\theta^{\sigma_r(k)+1})_{k\in \mathbb N}$ also converge to $\theta^*$. 

For $k$ sufficiently large, we may assume that the terms
$(\theta^{\sigma_r(k)+1},\theta^{\sigma_r(k)})$ and $(\theta,\theta^{\sigma_r(k)})$ belong to a compact neighborhood $C^*$
of $(\theta^*,\theta^*)$ included in $D_I$. By Definition \ref{SpAlt} of the Space Alternating Penalized Kullback Proximal iterations, 
$$
F_{\beta_{\sigma_r(k)}}(\theta^{\sigma_r(k)+1},\theta^{\sigma_r(k)}) \geq
F_{\beta_{\sigma_r(k)}}(\theta,\theta^{\sigma_r(k)}).
$$
Therefore,
\begin{equation}
\label{eq0}
\begin{array}{rl}
F_{\beta^*}(\theta^{\sigma_r(k)+1},\theta^{\sigma_r(k)})
& -(\beta_{\sigma_r(k)}-\beta^*) I_y(\theta^{\sigma_r(k)+1},\theta^{\sigma_r(k)}) \geq \\ 
&  F_{\beta^*}(\theta,\theta^{\sigma_r(k)})-(\beta_{\sigma_r(k)}-\beta^*)
I_y(\theta,\theta^{\sigma(k)}).
\end{array}
\end{equation}
Continuity of $F_{\beta}$ follows directly from the proof of Theorem 3.2.1 in Chr\'etien and Hero (2008), where  
in that proof $\sigma(k)$ has to be replaced by $\sigma_r(k)$. This implies that 
\begin{equation}
\label{xbz}
F_{\beta^*}(\theta^*,\theta^*)\geq F_{\beta^*}(\theta,\theta^*) 
\end{equation}
for all $\theta\in \Theta_r$ such that $(\theta,\theta^*)\in C^*\cap D_I$. Finally, recall that no assumption was made on $\theta$, and that
$C^*$ is a compact neighborhood of $\theta^*$. Thus, using the
assumption \ref{ass1}(i), which asserts that $l_y(\theta)$ tends to
$-\infty$ as $\|\theta\|$ tends to $+\infty$, we may deduce that
(\ref{xbz}) holds for any $\theta \in \Theta_r$ such that $(\theta,\theta^*)
\in D_I$ and, letting $\epsilon$ tend to zero, we see that
$\theta^*$ maximizes $F_{\beta^*}(\theta,\theta^*)$ for all
$\theta \in \Theta_r$ such that $(\theta,\theta^*)$ belongs to $D_I$ as claimed.

To conclude the proof of Theorem \ref{bord}, take $d$ in $\mathbb R^p$ and decompose $d$ as $d=d_1+\cdots+d_R$ with $d_r\in \mathcal S_r$. Then, equation (\ref{xbz}) implies that the directional derivatives satisfy
\begin{equation}
\label{posdirdif}
F_{\beta^*}^\prime (\theta^*,\theta^*;d_r)\leq 0 
\end{equation}
for all $r=1,\ldots,R$. Due to Assumption \ref{ass1} (iv), the directional derivative of $\sum_{r=1}^R \lambda_r p_n(\psi_r(\cdot))$ in the direction $d$ 
is equal to the sum of the partial derivatives in the directions $d_1,\ldots,d_R$ and, since all other terms in the definition of 
$F_\beta$ are differentiable, we obtain using (\ref{posdirdif}), that  
\begin{equation}
F_{\beta^*}^\prime (\theta^*,\theta^*;d)=\sum_{r=1}^R F_{\beta^*}^\prime (\theta^*,\theta^*;d_r)\leq 0. 
\nonumber
\end{equation}
Therefore, using the assumption that $p_n$ is regular (see Asssumption 1(i)) which says that $p_n^\circ=p_n^\prime$, 
together with characterization (\ref{carac}) of the subdifferential in the Appendix and Proposition 
2.1.5 (a) in \cite{Clarke90}, the desired result follows. 
\hfill$\Box$

Next, we consider the case where Assumptions \ref{ass3} hold. 
\begin{theorem}
\label{bordp}
Assume that in addition to Assumptions \ref{ass1} and \ref{ass4}, Assumptions \ref{ass3} hold.
Let $\theta^*$ be a cluster point of the Space Alternating Penalized Kullback Proximal sequence. Assume that all the functions $t_{ij}$ are 
continuously differentiable at $\theta^*$. Let $\mathcal I^*$ denote the index of the active constraints at $\theta^*$, i.e. 
$\mathcal I^*=\{(i,j) \textrm{ s.t. } t_{ij}(\theta^*)=0\}$. 
If $\theta^*$ lies in the interior of $D_l$, then $\theta^*$ satisfies the following property: there
exists a family of subsets $\mathcal I_r^{**}\subset \mathcal I^*$ and a set of real numbers $\lambda_{ij}^*$, $(i,j)\in \mathcal I_r^{**}$, 
$r=1,\ldots,R$ such that
\begin{equation}
\label{weakkkt}
0\in  \nabla l_y(\theta^*)-\sum_{r=1}^R\lambda_r \partial p_n(\psi_r(\theta^*))
+\sum_{r=1}^R \sum_{(i,j)\in \mathcal I_r^{**}} \lambda_{ij}^*{\rm P_{\mathcal S_r}}(\nabla t_{ij}(\theta^*)),
\end{equation}
where ${\rm P_{\mathcal S_r}}$ is the projection onto $\mathcal S_r$. 
\end{theorem}

\begin{rem}
The condition (\ref{weakkkt}) resembles the traditional Karush-Kuhn-Tucker conditions of optimality but is in fact weaker since the vector 
$$\sum_{r=1}^R \sum_{(i,j)\in \mathcal I_r^{**}} \lambda_{ij}^*{\rm P_{\mathcal S_r}}(\nabla t_{ij}(\theta^*))$$ 
in equation (\ref{weakkkt}) does not necessarily belong to the normal
cone at $\theta^*$ to the set $\{\theta \mid t_{ij}\geq 0,\: i=1,\ldots,n,\: j=1,\ldots,m \}$.
\end{rem}

{\bf Proof of Theorem \ref{bordp}}. Let $\Phi_{ij}(\theta,\bar{\theta})$ denote the bivariate function defined by
$$\Phi_{ij}(\theta,\bar{\theta})= \phi\Big(\frac{t_{ij}(\bar{\theta})}{t_{ij}(\theta)}\Big).$$

As in the proof of Theorem \ref{bord}, let $(x^{\sigma(k)})_{k\in \mathbb N}$ be a subsequence of iterates of (\ref{kullprox}) converging to $\theta^*$. Moreover let $r=1,\ldots,R$ and 
$\theta \in \Theta_r\cap D_l$. For each $k$, let $\sigma_r(k)$ be the next index greater than $\sigma(k)$ such that $(\sigma_r(k)-1) \:({\rm mod}\: R)+1=r$. 
Using the fact that $t$ is injective on every $\Theta_r$, $r=1,\ldots,R$, Lemma \ref{yal} and the fact that $(\beta_k)_{k\in \mathbb N}$ converges to $\beta^*>0$, we easily conclude that $(\theta^{\sigma_r(k)})_{k\in \mathbb N}$ and $(\theta^{\sigma_r(k)+1})_{k\in \mathbb N}$ also converge to $\theta^*$. 

Due to Assumption \ref{ass3} (iv), the first order optimality condition at iteration $\sigma_r(k)$ can be written
\begin{equation}
\label{frst}
\begin{array}{rl}
0=  & {\rm P_{\mathcal S_r}}(\nabla l_y(\theta^{\sigma(k)+1})) - \lambda_r g_r^{\sigma_r(k)+1} +\beta_{\sigma_r(k)}\Big(\sum_{ij} \alpha_{ij}(y_j)
{\rm P_{\mathcal S_r}}(\nabla t_{ij}(\theta^{\sigma_r(k)+1}))\\
& \Phi_{ij}(\theta^{\sigma_r(k)+1},\theta^{\sigma_r(k)}) 
 +\sum_{ij} \alpha_{ij}(y_j) t_{ij}(\theta^{\sigma_r(k)+1})
{\rm P_{\mathcal S_r}}(\nabla_1 \Phi_{ij}(\theta^{\sigma_r(k)+1},\theta^{\sigma_r(k)}))\Big) 
\end{array}
\end{equation}
with $g_r^{\sigma_r(k)+1}\in \partial p_n(\psi_r(\theta^{\sigma_r(k)+1}))$. 

Moreover, Claim A in the proof of Theorem 4.2.1 in Chr\'etien and Hero (2008), gives that for all $(i,j)$ such that $\alpha_{ij}(y_j)\neq 0$
\begin{equation}
\label{claimA}
\lim_{k\rightarrow +\infty}t_{ij}(\theta^{\sigma_r(k)+1})\nabla_1 \Phi_{ij}(\theta^{\sigma_r(k)+1},\theta^{\sigma_r(k)})=0.
\end{equation}
Let $\mathcal I_r^{*}$ be a subset of indices such that the family $\{{\rm P_{\mathcal S_r}}(\nabla t_{ij}(\theta^*))\}_{(i,j)\in \mathcal I_r^{*}}$ is linearly independent and spans the linear space generated by the family of all projected gradients 
$\{{\rm P_{\mathcal S_r}}(\nabla t_{ij}(\theta^*))\}_{i=1,\ldots,n,j=1,\ldots,m}$. 
Since this linear independence are preserved under small perturbations (continuity of the gradients), 
we may assume, without loss of generality, that the family $$\Big\{{\rm P_{\mathcal S_r}}(\nabla t_{ij}(\theta^{\sigma_r(k)+1}))\Big\}_{(i,j)\in \mathcal I_r^{*}}$$ is linearly independent
for $k$ sufficiently large. For such $k$, we may thus rewrite equation (\ref{frst}) as 
\begin{equation}
\label{frstbis}
\begin{array}{rl}
0= &{\rm P_{\mathcal S_r}}(\nabla l_y(\theta^{\sigma_r(k)+1}))  - \lambda_r g_r^{\sigma_r(k)+1} +\beta_{\sigma_r(k)}\Big(\sum_{(i,j)\in \mathcal I_r^{*}} \pi^{\sigma_r(k)+1}_{ij}(y_j) \\
& {\rm P_{\mathcal S_r}}(\nabla t_{ij}(\theta^{\sigma_r(k)+1}))+\sum_{ij} \alpha_{ij}(y_j) t_{ij}(\theta^{\sigma_r(k)+1})
{\rm P_{\mathcal S_r}}(\nabla_1 \Phi(\theta^{\sigma_r(k)+1},\theta^{\sigma_r(k)}))\Big), 
\end{array}
\end{equation}
where 
\begin{equation}
\label{paille}
\pi_{ij}^{\sigma_r(k)+1}(y_j)=\alpha_{ij}(y_j) \Phi_{ij}(\theta^{\sigma_r(k)+1},\theta^{\sigma_r(k)}).
\end{equation}

{\bf Claim}. {\em The sequence $\{\pi^{\sigma_r(k)+1}_{ij}(y_j)\}_{k\in \mathbb N}$ has a convergent subsequence for all $(i,j)$ 
in $I_r^{*}$.}

{\bf Proof of the claim}. Since the sequence $(\theta^k)_{k\in \mathbb N}$ is bounded, $\psi$ is continuously differentiable 
and the penalty $p_n$ has bounded subdifferential on compact sets, there exists a convergent subsequence 
$(g_r^{\sigma_r(\gamma(k))+1})_{k\in \mathbb N}$ with limit $g_r^*$. Now, using Equation (\ref{claimA}), this last equation 
implies that $\{\pi^{\sigma_r(\gamma(k))+1}_{(i,j)\in \mathcal I_r^{*}}(y_j)\}_{(i,j)\in \mathcal I_r^{*}}$ converges to the
coordinates of a vector in the linearly independent family 
$\{{\rm P_{\mathcal S_r}}(\nabla t_{ij}(\theta^{*}))\}_{(i,j)\in \mathcal I_r^{*}}$. This concludes the proof. \hfill $\Box$

The above claim allows us to finish the proof of Theorem \ref{bordp}. Since a subsequence
$(\pi^{\sigma_r(\gamma(k))+1}_{ij}(y_j))_{(i,j)\in \mathcal I_r^{*}}$ is convergent, we may consider its limit 
$(\pi^*_{ij})_{(i,j)\in \mathcal I_r^{*}}$. Passing to the limit, we obtain from equation (\ref{frst}) that
\begin{equation}
\label{clustr}
0={\rm P_{\mathcal S_r}}(\nabla l_y(\theta^*))-\lambda_r g_r^*+\beta^*\Big(\sum_{(i,j)\in \mathcal I_r^{*}} \pi_{ij}^* 
{\rm P_{\mathcal S_r}}(\nabla t_{ij}(\theta^*))\Big).
\end{equation}
Using the outer semi-continuity property of 
the subdifferential of locally Lipschitz functions (see Appendix) we thus obtain that $g_r^*\in \partial p_n(\psi_r(\theta^*))$. 
Now, summing over $r$ in (\ref{clustr}), we obtain
$$
0= \sum_{r=1}^R {\rm P_{\mathcal S_r}}(\nabla l_y(\theta^*))-\sum_{r=1}^R\lambda_r g^*_r+\beta^*\sum_{r=1}^R\Big(\sum_{(i,j)\in \mathcal I_r^{*}} \pi_{ij}^* {\rm P_{\mathcal S_r}}(\nabla t_{ij}(\theta^*))\Big).
$$
Moreover, since $\Phi_{ij}(\theta^{\sigma_r(k)+1},\theta^{\sigma_r(k)})$ tends to zero if $(i,j)\not\in \mathcal I^*$, i.e. if the constraint on 
component $(i,j)$ is not active, equation (\ref{paille}) implies that
$$
0= \sum_{r=1}^R {\rm P_{\mathcal S_r}}(\nabla l_y(\theta^*))-\sum_{r=1}^R\lambda_r g^*_r+\beta^*\sum_{r=1}^R\Big(\sum_{(i,j)\in \mathcal I_r^{**}} \pi_{ij}^* {\rm P_{\mathcal S_r}}(\nabla t_{ij}(\theta^*))\Big)
$$
where $\mathcal I_r^{**}$ is the subset of active indices of $\mathcal I_r^{*}$, i.e. $\mathcal I_r^{**}=\mathcal I_r^*\cap \mathcal I^*$. 
Since $\sum_{r=1}^R\lambda_r g^*_r \in \sum_{r=1}^R\lambda_r \partial p_n(\psi_r(\theta^*))$, this implies that  
\begin{equation}
\label{prekkt}
0\in  \nabla l_y(\theta^*)-\sum_{r=1}^R\lambda_r \partial p_n(\psi_r(\theta^*))
+\beta^*\sum_{r=1}^R \sum_{(i,j)\in \mathcal I_r^{**}} \pi_{ij}^*{\rm P_{\mathcal S_r}}(\nabla t_{ij}(\theta^*)),
\end{equation}
which establishes Theorem \ref{bordp} once we define $\lambda_{ij}^*=\lambda^*\pi_{ij}^*$.
\hfill$\Box$

The result (\ref{prekkt}) can be refined to the classical Karush-Kuhn-Tucker type condition under additional conditions such as stated below. 
\begin{corollary}
\label{coro}
If in addition to the assumptions of Theorem \ref{bordp} we assume that either ${\rm P}_{\mathcal S_r}(\nabla t_{ij}(\theta^*))=\nabla t_{ij}(\theta^*)$ or ${\rm P}_{\mathcal S_r}(\nabla t_{ij}(\theta^*))=0$ for all $(i,j)\in \mathcal I^*$, i.e. such that $t_{ij}(\theta^*)=0$, then there
exists a set of subsets $\mathcal I_r^{**}\subset \mathcal I^*$ and a family of real numbers $\lambda_{ij}^*$, $(i,j)\in \mathcal I_r^{**}$ , $r=1,\ldots,R$ such that the following Karush-Kuhn-Tucker condition for optimality holds at cluster point $\theta^*$: 
$$
0\in  \nabla l_y(\theta^*)-\sum_{r=1}^R\lambda_r \partial p_n(\psi_r(\theta^*))
+\sum_{r=1}^R \sum_{(i,j)\in \mathcal I_r^{**}} \lambda_{ij}^*\nabla t_{ij}(\theta^*).
$$
\end{corollary}

\section{Application: Variable selection in finite mixtures of regression models}
\label{mixreg}

Variable subset selection in regression models is frequently performed using penalization of the likelihood function, 
e.g. using AIC, Akaike (1973) and BIC, Schwarz (1978) penalties. 
The main drawback of these approaches is lack of scalability due to a combinatorial explosion of the 
set of possible models as the number of variables increases. 
Newer methods use $l_1$-type penalties of likelihood functions, as in the LASSO, Tibshirani (1996) and the Dantzig selector of Cand\`es and Tao 
(2007), to select subsets of variables without enumeration. 

Computation of maximizers of the penalized likelihood function 
can be performed using standard algorithms for nondifferentiable optimization such as 
bundle methods, as introduced in Hiriart-Urruty and Lemar\'echal (1993). However general purpose optimization methods might be difficult 
to implement in the situation where, for instance, log objective functions induce line-search problems. In certain cases, the EM algorithm, 
or a combination of EM type methods with general purpose optimization routines might be simpler to implement. Variable 
selection in finite mixture models, as described in Khalili and Chen (2007), represents such a case due to the presence of very natural hidden variables.

In the finite mixture estimation problem considered here, $y_1,\ldots,y_n$ are realizations of the response variable $Y$ and $x_1,\ldots,x_n$ 
are the associated realizations of the $P$-dimensional vector of covariates $X$. We focus on the case of a mixture of linear regression models sharing the 
same variance, as in the baseball data example of section 7.2 in Khalili and Chen (2007), i.e. 
\begin{equation}
Y\sim \sum_{k=1}^K \pi_k \mathcal N(X^t \beta_k,\sigma^2),
\end{equation}
with $\pi_1,\ldots,\pi_k \geq 0$ and $\sum_{k=1}^K\pi_k=1$. The main problem discussed in Khalili and Chen (2007) is model selection for 
which a generalization of the smoothly clipped absolute deviation (SCAD) method of Fan and Li (2001,2002) is proposed using an MM-EM algorithm in the spirit of 
Hunter and Lange (2004). No convergence property of the MM algorithm was established. The purpose of this section is to show that the Space Alternating KPP EM generalization 
is easily implemented and that stationarity of the cluster points is garanteed by the theoretical analysis of Section \ref{theo}. 

The SCAD penalty, studied in Khalili and Chen (2007) is a modification of the $l_1$ penalty which is given by 
\begin{equation}
p_n(\beta_1,\ldots,\beta_K)=\sum_{k=1}^K \pi_k \sum_{j=1}^P p_{\gamma_{nk}}(\beta_{k,j})  
\nonumber
\end{equation}
where $p_{nk}$ is specified by 
\begin{equation}
p_{\gamma_{nk}}^\prime(\beta)=\gamma_{nk}\sqrt{n} 1_{\sqrt{n}|\beta|\leq \gamma_{nk}}+\frac{\sqrt{n}(a\gamma_{nk}-\sqrt{n}|\beta|)_+}{a-1}1_{\sqrt{n}|\beta|>\gamma_{nk}}
\nonumber
\end{equation}
for $\beta$ in $\mathbb R$. 

Define the missing data as the class labels $z_1,\ldots,z_n$ of the mixture component from which the observed data point $y_n$ was drawn. 
The complete log-likelihood is then 
\begin{equation}
l_c(\beta_1,\ldots,\beta_K,\sigma^2)=\sum_{i=1}^n \log(\pi_{z_{i}})- \frac12\log (2\pi \sigma^2)-\frac{(y_i-x_i^t\beta_{z_{i}})^2}{2\sigma^2}.
\nonumber
\end{equation}
Setting $\theta=(\pi_1,\ldots,\pi_K,\beta_1,\ldots,\beta_K,\sigma^2)$, the penalized $Q$-function is given by 
\begin{equation}
Q(\theta,\bar{\theta})=\sum_{i=1}^n \sum_{k=1}^K t_{ik}(\bar{\theta}) \left[\log(\pi_k)- \frac12\log (2\pi \sigma^2)-\frac{(y_i-x_i^t\beta_k)^2}{2\sigma^2} \right]-p_n(\beta_1,\ldots,\beta_K)
\nonumber
\end{equation}
where 
\begin{equation}
t_{ik}(\theta)=\frac{\pi_k\frac1{\sqrt{2\pi\sigma^2}} \exp \Big(-\frac{(y_i-X\beta_k)^2}{2\sigma^2}\Big)}
{\sum_{l=1}^K\pi_l\frac1{\sqrt{2\pi\sigma^2}} \exp \Big(-\frac{(y_i-X\beta_l)^2}{2\sigma^2}\Big)}.
\nonumber
\end{equation}

The computation of this $Q$-function accomplishes the E-step. Moreover, a penalty of the form 
$-\sum_{k=1}^K\sum_{j=1}^P |\max\{10^6,|\beta_{k,j}|\}-10^6|$ can be added to 
the log-likelihood function in order to ensure that Assumptions 1(i) (convergence of the penalized log-likelihood to $-\infty$ for 
parameter values with norm growing to $+\infty$) is satisfied for the case where $X$ is not invertible. 
Due to the fact that the penalty $p_n$ is a function of the mixture probabilities $\pi_k$, the M-step estimate of the 
$\pi$ vector is not given by the usual formula 
\begin{equation}
\label{approxPiupdates}
\pi_k=\frac1{n} \sum_{i=1}^n t_{ik}(\bar{\theta}) \hspace{.4cm} k=1,\ldots,K.
\end{equation}
This, however, is the choice made in Khalili and Chen (2007) in their implementation. 
Moreover, optimizing jointly over the variables $\beta_k$ and $\pi_k$ is clearly a more complicated task than independently optimizing 
with respect to each variable. We implement a componentwise version of EM 
consisting of successively optimizing with respect to the $\pi_k$'s and 
alternatively with respect to the vectors $\beta_k$. Optimization with respect to the $\pi_k$'s can be easily performed using standard differentiable optimization routines 
and optimization with respect to the $\beta_k$'s can be performed by a standard non-differentiable optimization routine, e.g.
as provided by the function {\sf optim} of Scilab using the {\sf 'nd'} (standing for 'non-differentiable') option.  

We now turn to the description of the Kullback proximal penalty $I_y$ defined by (\ref{kullb}). The conditional density function $k(y_1,\ldots,y_n,z_1,\ldots,z_n\mid y_1,\ldots,y_n;\theta)$ is
$$
k(y_1,\ldots,y_n,z_1,\ldots,z_n\mid y_1,\ldots,y_n;\theta)=\prod_{i=1}^n t_{iz_i}(\theta).
$$
and therefore, the Kullback distance-like function $I_y(\theta,\bar{\theta})$ is
\begin{equation}
\label{KLmix}
I_y(\theta,\bar{\theta})= \sum_{i=1}^n \sum_{k=1}^K t_{ik}(\bar{\theta})\log\Big(\frac{t_{ik}(\bar{\theta})}{t_{ik}(\theta)} \Big).
\end{equation}

We have $R=K+1$ subsets of variables with respect to which optimization will be performed successively. All components of Assumptions \ref{ass1} and \ref{ass3} are trivially satisfied for this model. Validation of Assumption \ref{ass3} (iv)
is provided by Lemma 1 of Celeux {\em et al}. (2001). On the other hand, since $t_{ik}(\theta)=0$ implies that $\pi_k=0$
and $\pi_k=0$ implies
\begin{equation}
\frac{\partial t_{ik}}{\partial \beta_{jl}}(\theta)=0
\nonumber
\end{equation}
for all $j=1,\ldots,p$ and $l=1,\ldots,K$ and 
\begin{equation}
\frac{\partial t_{ik}}{\partial \sigma^2}(\theta)=0,
\nonumber
\end{equation}
it follows that $P_{{\mathcal S}_r}(\nabla t_{ik}(\theta^*))=\nabla t_{ik}(\theta^*)$ if $\mathcal S_r$ is the vector space generated by 
the probability vectors $\pi$ and $P_{{\mathcal S}_r}(\nabla t_{ik}(\theta^*))=0$ otherwise. 
Therefore, Corollary \ref{coro} applies. 

We illustrate this algorithm on real data (available at 

{\em http://www.amstat.org/publications/jse/v6n2/datasets.watnik.html}). 

\begin{figure}
  \includegraphics{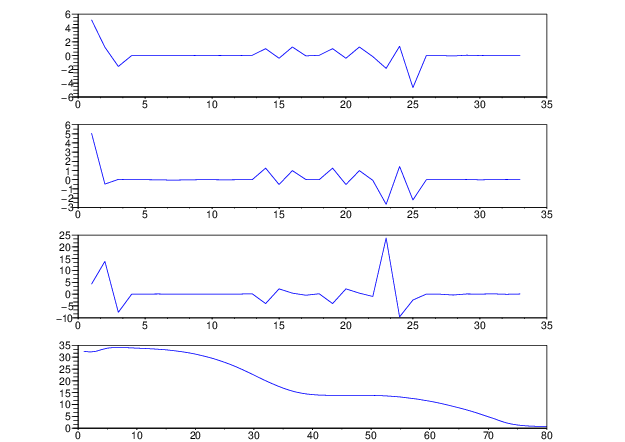}
\caption{Baseball data of Khalili and Chen (2007). This experiment is performed with the plain EM. The parameters are $\gamma_{nk}=.1$ and $a=10$. The first plot is the vector $\beta$ obtained for the single component model. The second (resp. third) plot is the vector of the optimal $\beta_1$ (resp. $\beta_2$). The fourth plot is the euclidean distance to the optimal $\theta^*$ versus iteration index. The starting value of $\pi_1$ was .3}
\label{fig1}       
\end{figure}

Khalili and Chen (2007) report that a model with only two components was selected by the BIC criterion in comparison to a three components
model. Here, two alternative algorithms are compared: the approximate EM using (\ref{approxPiupdates}) and the plain EM using the {\sf optim} subroutines. The 
results for $\gamma_{nk}=1$ and $a=10$ are given in Figures \ref{fig1}. 

\begin{figure}
  \includegraphics{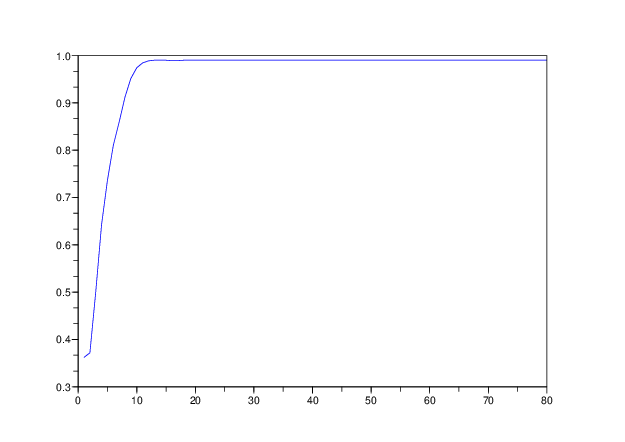}
\caption{This experiment is performed with the plain EM for the Baseball data of Khalili and Chen (2007). 
The parameters are $\gamma_{nk}=5$ and $a=10$. The plot shows the 
probability $\pi_1$ of the first component versus iteration index. The starting value of $\pi_1$ was .3}
\label{degpi}       
\end{figure}

\begin{figure}
  \includegraphics{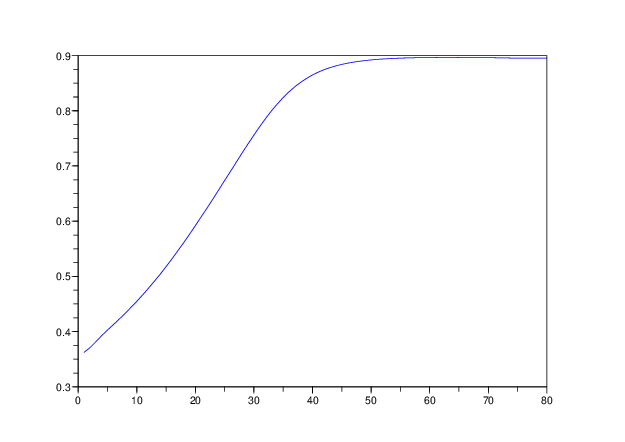}
\caption{Baseball data of Khalili and Chen (2007). This experiment is performed with the approximate EM. The parameters are $\gamma_{nk}=5$ and $a=10$. The plot shows the 
probability $\pi_1$ of the first component versus iteration index. The starting value of $\pi_1$ was .3}
\label{nondegpi}       
\end{figure}

The results shown in Figure \ref{fig1} establish that the approximate EM algorithm has similar properties to the plain EM algorithm for small values of the threshold parameters $\gamma_{nk}$. 
Moreover, the larger the values of $\gamma_{nk}$, the closer the probability of the first component is to 1. One important fact to notice is that with the plain 
EM algorithm, the optimal probability vector becomes singular, in the sense that the second component has zero probability, as shown in 
Figure \ref{degpi} . Figure \ref{nondegpi} demonstrates that 
the approximate EM algorithm of Khalili and Chen (2007) does not produce optimal solutions.  

\section{Conclusion}

In this paper we analyzed the expectation maximization (EM)
algorithm with non-differentiable penalty.
By casting the EM algorithm as a Kullback Proximal Penalized (KPP) iteration,
we  proved the stationarity of the cluster points and showed that
any cluster point of the Space Alternating KPP method satisfies a nonsmooth Karush-Kuhn-Tucker condition.
The theory was applied to a space alternating implementation of the
penalized EM algorithm for a problem of model selection in a finite mixture
of linear regressions.

\section{Appendix: The Clarke subdifferential of a locally Lipschitz function}
Since we are dealing with non differentiable functions, the notion of generalized 
differentiability is required. The main references for this appendix are Clarke (1990) and Rockafellar and Wets (2004). 
A locally Lipschitz function $f$: $\mathbb R^p\mapsto \mathbb R$ always has a generalized directional derivative 
$f^\circ(\theta,\omega)$: $\mathbb R^p\times \mathbb R^p\mapsto \mathbb R$ in the sense given by Clarke, i.e. 
\begin{equation}
f^\circ(\theta,\omega)={\textrm{lim sup}}_{\eta \in \mathbb R^p\rightarrow \theta,\: t\downarrow 0} \frac{f(\eta+t\omega)-f(\eta)}{t}. 
\nonumber
\end{equation}
A locally Lipschitz function is called {\em regular} if it admits a directional derivative at every point and 
if moreover this directional derivative coincides with Clarke's generalized directional derivative. 

The Clarke subdifferential of $f$ at $\theta$ is the convex set defined by 
\begin{equation}
\label{carac}
\partial f(\theta)=\{\eta \mid f^\circ(\theta,\omega)\geq \eta^t\omega, \:\: \forall \omega \}. 
\end{equation} 
\begin{proposition}
\label{single}
The function $f$ is differentiable if and 
only if $\partial f(\theta)$ is a singleton.  
\end{proposition}
We now introduce another very important property of the Clarke subdifferential related to generalization of semicontinuity for
set-valued maps.  
\begin{definition}
A set-valued map $\Phi$ is said to be outer-semicontinuous if its graph 
\begin{equation}
{\rm graph}\: \Phi =\{(\theta,g) \mid g\in \Phi(\theta) \}
\nonumber
\end{equation}
is closed, i.e. if for any sequence (${\rm graph} \Phi \ni$) $(\theta_n,g_n) \rightarrow (\theta^*,g^*)$  as $n\rightarrow +\infty$, then 
$(\theta^*,g^*)\in {\rm graph} \Phi$. 
\end{definition}
One crucial property of the Clarke subdifferential is that it is outer-semicontinuous. 

A point $\theta$ is said to be a {\em stationary point} of $f$ if 
\begin{equation}
0\in \partial f(\theta). 
\nonumber
\end{equation}
Consider now the problem
\begin{equation}
\sup_{\theta \in \mathbb R^p} f(\theta) 
\nonumber
\end{equation}
subject to 
\begin{equation}
g(\theta)=[g_1(\theta),\ldots,g_m(\theta)]^t\geq 0
\nonumber
\end{equation}
where all the functions are locally Lipschitz from $\mathbb R^p$ to $\mathbb R$. 
Then, a necessary condition for optimality of $\theta$ is the Karush-Kuhn-Tucker condition, i.e. there exists a vector $u\in \mathbb R_+^m$ such 
that 
\begin{equation}
0\in \partial f(\theta)+\sum_{j=1}^m u_j \partial g_j(\theta).
\nonumber
\end{equation}
Convex functions are in particular locally Lipschitz. The main references for these facts are Rockafellar (1970) and Hiriart-Urruty and Lemar\'echal (1993).



\end{document}